\documentclass[aps,prl,twocolumn,superscriptaddress,floatfix]{revtex4-1}
\usepackage{makecell}
\usepackage{graphicx}% Include figure files
\usepackage{dcolumn}% Align table columns on decimal point
\usepackage{bm}% bold math
\usepackage{times}% font that prx use
\usepackage{color}
\usepackage{appendix}
\usepackage{threeparttable}
\usepackage[colorlinks=true,linkcolor=blue,anchorcolor=blue,citecolor=blue,urlcolor=blue ]{hyperref}
\usepackage{gensymb}
\usepackage{ulem}
\usepackage{amssymb}
\usepackage{makecell}

\begin{document}
\bibliographystyle{apsrev4-1}
\title{Field-induced topological Hall effect and butterfly-shaped magnetoresistance in the centrosymmetric antiferromagnet EuAuAs}

\author{Yu Zhang}
\affiliation{School of Physics and Key Laboratory of Quantum State Construction and Manipulation (Ministry of Education), Renmin University of China, Beijing 100872, China}

\author{Junfa Lin}
\affiliation{School of Physics and Key Laboratory of Quantum State Construction and Manipulation (Ministry of Education), Renmin University of China, Beijing 100872, China}

\author{Huan Wang}
\affiliation{School of Physics and Key Laboratory of Quantum State Construction and Manipulation (Ministry of Education), Renmin University of China, Beijing 100872, China}
\affiliation{School of Police Equipment Technology, China People's Police University, Langfang 065000, Hebei, People's Republic of China}

\author{Kun Han}
\affiliation{School of Physics and Key Laboratory of Quantum State Construction and Manipulation (Ministry of Education), Renmin University of China, Beijing 100872, China}

\author{Yiting Wang}
\affiliation{School of Physics and Key Laboratory of Quantum State Construction and Manipulation (Ministry of Education), Renmin University of China, Beijing 100872, China}

\author{Xue Dong}
\affiliation{School of Physics and Key Laboratory of Quantum State Construction and Manipulation (Ministry of Education), Renmin University of China, Beijing 100872, China}

\author{Zhenfeng Guan}
\affiliation{School of Physics and Key Laboratory of Quantum State Construction and Manipulation (Ministry of Education), Renmin University of China, Beijing 100872, China}

\author{Shengdi Xi}
\affiliation{School of Physics and Key Laboratory of Quantum State Construction and Manipulation (Ministry of Education), Renmin University of China, Beijing 100872, China}

\author{Tian-Long Xia}\email{tlxia@ruc.edu.cn}
\affiliation{School of Physics and Key Laboratory of Quantum State Construction and Manipulation (Ministry of Education), Renmin University of China, Beijing 100872, China}
\affiliation{ Laboratory for Neutron Scattering, Renmin University of China, Beijing 100872, People's Republic of China}

\date{\today}

\begin{abstract}
 The coupling between magnetic and electronic degrees of freedom gives rise to a variety of intriguing transport phenomena. Among them, the topological Hall effect, originating from the real-space Berry phase associated with nontrivial magnetic textures, has attracted considerable attention. Here, we systematically investigate the magnetic and transport properties of antiferromagnet EuAuAs. Magnetic characterizations reveal antiferromagnetic transition at 5.7 K and 6.3 K for $H \parallel ab$ and $H \parallel c$,  accompanied by metamagnetic transition and small hysteresis for $H \parallel ab$. Electrical transport measurements reveal a pronounced topological Hall effct in the antiferromagnetic state with $H \parallel ab$ and $I \parallel c$, which may be attributed to finite scalar spin chirality. Furthermore, the magnetoresistance exhibits butterfly-shaped hysteresis and strong angular dependence, which are likely associated with spin-dependent electron scattering, magnetic-domain evolution, and domain-wall pinning. Our results suggest that field-induced spin textures play an important role in the magnetotransport properties and provide insights into the interplay between magnetic textures and electronic transport in centrosymmetric antiferromagnets.

\end{abstract}

\maketitle
\section{\uppercase\expandafter{\romannumeral1}. Introduction}
Transport phenomena in magnetic materials have been intensively investigated, particularly Hall responses associated with complex magnetic structures, such as the anomalous Hall effect (AHE) and the topological Hall effect (THE). Initially, the AHE manifests as a transverse voltage proportional to the magnetization in ferromagnetic systems,  even in the absence of an external magnetic field \citep{RevModPhys.82.1539,PhysRev.95.1154,PhysRevLett.96.037204}. Recent studies have revealed that the AHE can be observed in noncollinear antiferromagnets such as 
Mn$_3$Sn \citep{10.1038/nature15723}, Mn$_3$Ge \citep{doi:10.1126/sciadv.1501870}. To date, the origin of the AHE has been extensively studied and is generally attributed to two main mechanisms: One is the extrinsic mechanism, which arises from impurity scattering , including skew scattering and side-jump \citep{PhysRevB.2.4559,10.1016/S0031-8914(55)92596-9,10.1016/S0031-8914(58)93541-9}. The other is the intrinsic Karplus–Luttinger mechanism, originating from the Berry curvature in momentum space \citep{10.1126/science.1089408,RevModPhys.82.1959}. In contrast to the intrinsic AHE determined by the Berry curvature in momentum space, the topological Hall effect (THE) originates from the real-space Berry curvatures induced by chiral spin textures \citep{PhysRevLett.93.096806,10.1038/s43246-022-00238-2,https://doi.org/10.1016/j.pmatsci.2022.100971}. When conduction electrons propagate through a noncoplanar spin configuration, they acquire a nozero Berry phase that acts as an effective magnetic field in real space. Such spin textures possesses a finite scalar spin chirality (SSC), defined as $\boldsymbol	{\chi}_{ijk}$ = $\textbf{S}_{i} \cdot (\textbf{S}_{j}\times \textbf{S}_{k}$), where $\textbf{S}_{i}$, $\textbf{S}_{j}$ and $\textbf{S}_{k}$ represent spins at three neighboring atomic sites ($i$, $j$, $k$) \citep{PhysRevB.39.11413,RevModPhys.82.1539,10.1126/science.1058161}.
A nonzero SSC can arise in various noncoplanar magnetic configurations, including skyrmion textures \citep{PhysRevLett.106.156603}, field-induced spin-reorientation states associated with a spin-flop transition \citep{PhysRevB.100.024434}, and short-range spin clusters generated by thermal fluctuations and magnetic fields \citep{10.1002/adfm.202502016,10.1126/sciadv.aap9962,PhysRevLett.108.056601}. Early studies focused primarily on noncentrosymmetric B20-type compounds, including MnSi \citep{PhysRevLett.102.186602}, MnGe \citep{PhysRevLett.106.156603}, and FeGe thin films \citep{PhysRevLett.108.267201}, which reveal that magnetic skyrmionic topological spin textures can induce the THE. Moreover, the THE has also been observed in a centrosymmetric system, like the geometrically frustrated triangular‑lattice magnet Gd$_{2}$PdSi$_{3}$ \citep{10.1126/science.aau0968}. The application of a magnetic field induces a Bloch‑type skyrmion lattice accompanied by a giant THE, demonstrating that geometric frustration alone can stabilize topological spin states in centrosymmetric materials. Other centrosymmetric compounds, such as the kagome ferromagnet Fe$_{3}$Sn$_{2}$ \citep{10.1063/1.5088173}, MnNiGa\citep{10.1002/adma.201600889} display large THE over wide temperature ranges, which have been attributed to frustrated or noncoplanar spin textures.

Recently, centrosymmetric europium-based ternary materials of the general formula EuTX (T = transition metal, X = pnictogen) provide a fertile platform for investigating the novel physical phenomena. These compounds crystallize in the hexagonal ZrBeSi‑type structure and host divalent Eu$^{\text{2+}}$ ions with a half‑filled 4$f$ shell, which provides large localized moments. By tuning the T and X elements, they exhibit diverse magnetic ground states and topological properties. For instance, EuAgAs is an antiferromagnetic (AFM) Dirac semimetal exhibiting a large THE and chiral anomaly induced positive longitudinal magnetoconductivity \citep{PhysRevB.103.L241112},
EuCuAs displays a field-tunable THE associated with a field-induced transition from a helical to a transverse conical magnetic structure \citep{10.1021/jacs.3c04249}, and the centrosymmetric AFM EuCuSb exhibits a significant THE in the field range between the metamagnetic transition field and magnetic saturation field \citep{PhysRevB.111.054428}. Table I summarizes the reported magnetic ground states and Hall responses in representative EuTX compounds. Within this family, EuAuAs has been identified as an antiferromagnetic topological nodal-line semimetal candidate \citep{PhysRevB.105.045103}. However, a detailed understanding about magnetism and transport properties still remain unclear, and further investigation is needed to learn the relation between magnetic properties and electronic transport.

In this work, we successfully synthesized EuAuAs single crystals and systematically investigated magnetic and electronic transport properties. Magnetic measurements reveal that EuAuAs undergoes a antiferromagnetic transition at 5.7 K and 6.3 K for $H \parallel ab$ and $H \parallel c$. The $M(H)$ curves exhibit pronounced magnetic anisotropy between $H \parallel ab$ and $H \parallel c$. A distinct metamagnetic transition is observed for $H \parallel ab$, whereas the corresponding transition is absent for $H \parallel c$. Furthermore, a THE is observed and reaches a maximum value of approximately 0.35 $\mu\Omega$ cm for $H \parallel ab$ and $I \parallel c$ at 2 K. Our magnetic and electronic transport analyses reveal that the THE may originate from nonzero scalar spin chirality, which may be induced by a noncoplanar spin structure associated with the metamagnetic transition under $H \parallel ab$. Moreover, a weak butterfly-shaped magnetoresistance (MR) is observed, consistent with the field-dependent $M(H)$ curves. Notably, despite the smooth evolution of the $M(H)$ curves, the butterfly-shaped negative MR becomes more pronounced for $H \parallel c$, which may be due to AFM-domain existence and domain-wall pinning.

\begin{table}[htbp]
	\centering
	\caption{Different properties of EuTX (T = transition metal, X = pnictogen) materials. }
	\begin{threeparttable}
	\label{1}
	\renewcommand\arraystretch{1.5}
	\setlength{\tabcolsep}{0.05mm}{
		\begin{tabular}{cccccccc}
			\hline \hline 
			&  Compounds  &  Ground state  & AHE/THE& Ref. \\
			\hline
			&EuAuAs  &AFM &THE  & \ \ This work  \\
			&EuCuAs   &Helical magnetic&THE  & \citep{10.1021/jacs.3c04249}  \\
			&EuAgAs   &AFM  &THE  & \citep{PhysRevB.103.L241112} \\
			&EuAuSb	  &AFM   &THE  & \citep{PhysRevB.110.085145,PhysRevB.109.155152} \\
			&EuAgSb   &AFM  &AHE  & \citep{10.1016/j.jallcom.2024.178172}  \\
			&EuCuSb  &\makecell[c] {In-plane up-up-down-down \\  AFM$^{1}$  \& IC phase$^{2}$} &THE &\citep{PhysRevB.102.174425,PhysRevB.111.054428}    \\
			&EuAuBi    &Canted AFM  &THE  & \citep{dt26-56xc,arXiv:2604.25365}  \\
			&EuAgBi   &AFM  &Unknown  & \citep{10.59717/j.xinn-mater.2025.100171}   \\
			&EuCuBi   &AFM  &Unknown  &  \citep{PhysRevB.108.115126} \\
			&EuAuP    &FM &Unknown & \citep{10.1088/0953-8984/6/9/018,10.1016/j.commt.2024.100022}   \\
			&EuAgP   &FM   &Unknown  & \citep{/10.1016/j.jmat.2024.02.012,10.1016/j.mtphys.2021.100570}    \\
			&EuCuP   &FM & AHE & \citep{10.1016/j.jallcom.2023.169620,PhysRevMaterials.8.094202}     \\
			\hline 
	\end{tabular}}
\begin{tablenotes}
	\footnotesize
	\item[1] In-plane up-up-down-down AFM: Collinear antiferromagnetic with a spin configuration of up-up-down-down along the $c$ axis.
	\item[2] IC phase: The incommensurate helimagnetic phase.
\end{tablenotes}
	\end{threeparttable}
\end{table}

\begin{figure}[htbp]
	\centering
	\includegraphics[width=0.5\textwidth]{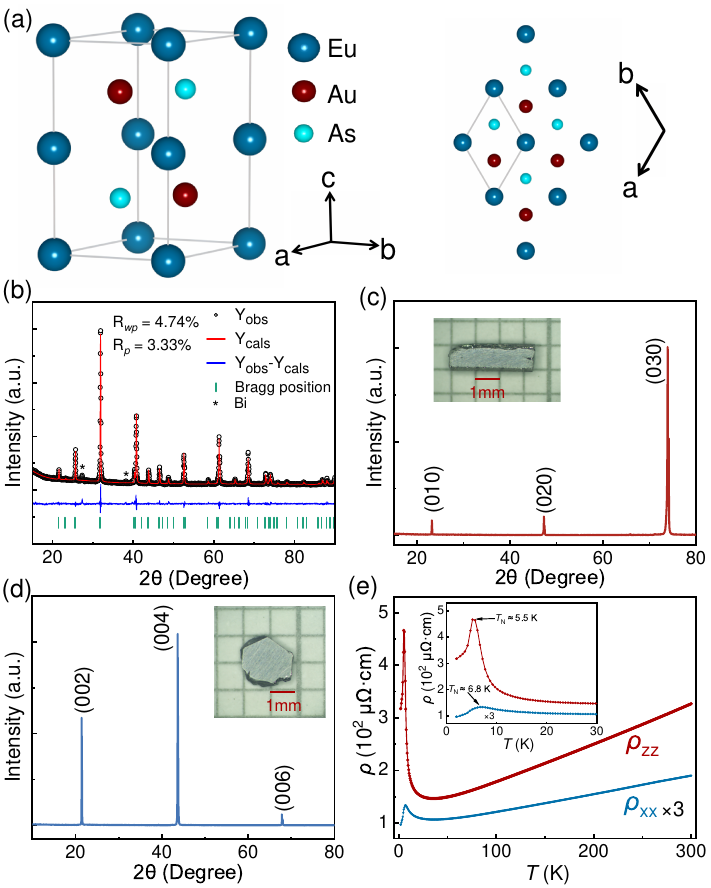}
	\caption{Crystal structure and longitudinal resistivity of EuAuAs. (a) Crystal structure of EuAuAs with space group P6$_{3}$/mmc (No.194). The Eu layers alternate with the Au–As layers along the c axis. (b) Room-temperature powder X-ray diffraction (XRD) pattern of ground EuAuAs single crystals and the corresponding Rietveld refinement. (c) and (d) XRD patterns of the single crystal with (0l0) Bragg peaks and (00l) Bragg peaks, respectively. Insets show the photographs of single crystals. (e) Temperature dependence of longitudinal resistivity $\rho_{xx}$ and $\rho_{zz}$ at zero field for $I \parallel ab$ and $I \parallel c$ , respectively. Inset, zoom in of the AFM transition temperatures with $T_{\text{N}}$ $\approx$ 5.5 K and $T_{\text{N}}$ $\approx$ 6.8 K for $I \parallel ab$ and $I \parallel c$. }  
	\label{Fig1} 	
\end{figure}

\begin{figure*}[htbp]
	\centering
	\includegraphics[width=1\textwidth]{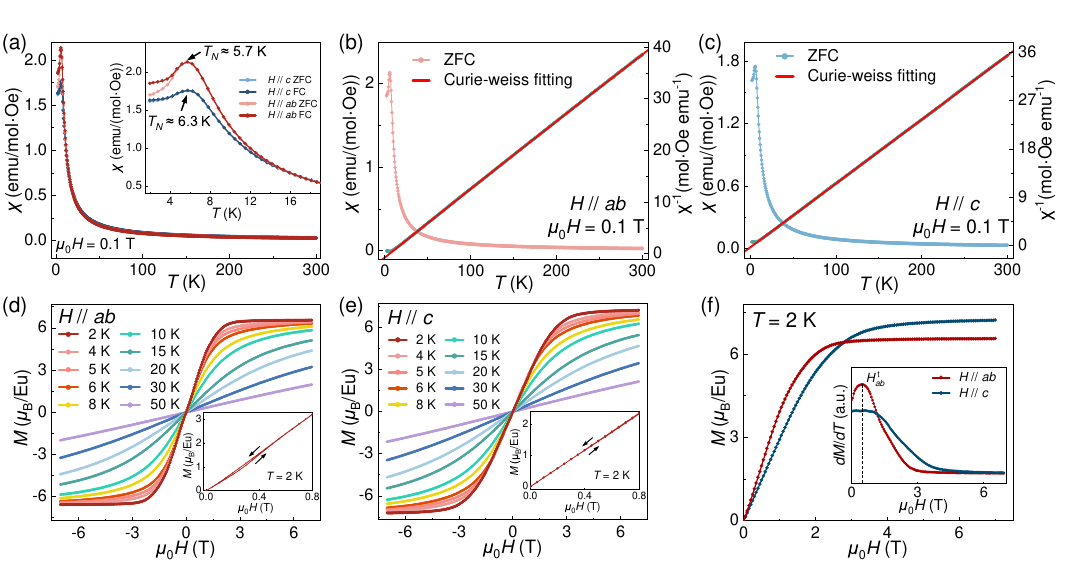}
	\caption{Magnetic properties of EuAuAs single crystal. (a) The temperature dependent magnetic susceptibility at zero-field cooled (ZFC) and field cooled (FC) with $\mu_{0}H$ = 0.1 T applied along the $c$-axis (blue line) and the $ab$-plane (red line), respectively. Inset, zoom in of the AFM transition temperatures with $T_{\text{N}}$ $\approx$ 5.7 K and $T_{\text{N}}$ $\approx$ 6.3 K for $H \parallel ab$ and $H \parallel c$.  (b) and (c) ZFC magnetic susceptibility $\chi$ and its inverse 1/$\chi$ as the function of temperature under $\mu_{0}H$ = 0.1 T . The field is applied along the $ab$‑plane in (b) and along the $c$‑axis in (c). The red lines represent the corresponding modified Curie-Weiss law fitting curves. (d) and (e) Field‑dependent magnetization curves  at various temperatures with $H \parallel ab$ and $H \parallel c$, respectively. The insets show the low‑field region ($\mu_{\text{0}}H$ $<$ 0.8 T) at $T$ = 2 K. (f) $M$–$H$ curves at 2 K for $H \parallel ab$ and $H \parallel c$ over the field range of 0 – 7 T. The inset shows the first derivative d$M$ / d$H$, where $H_{ab}^{\text{t}}$ is metamagnetic transition field for $H \parallel ab$.}  
	\label{Figure} 	
\end{figure*}

\section{\uppercase\expandafter{\romannumeral2}. Experimental Methods }

EuAuAs single crystals were grown using bismuth (Bi) flux. High-purity primary materials in a molar ratio of Eu : Au : As : Bi = 1:1:1:15 were mixed and put in an alumina crucible. The alumina crucible was sealed into an evacuated quartz ampoule and placed into a high-temperature muffle furnace, heated to 1050$\celsius$ and held for 20 h. The muffle furnace slowly cooled to 600$\celsius$ over 7 days. In order to obtain EuAuAs single crystals, the quartz tube was quickly removed  and inverted into the centrifuge to separate the excess Bi flux at 600$\celsius$. The stick-like single crystals were obtained and shown in the inset of Fig. 1(c) and Fig. 1(d). The chemical compisitions were confirmed by energy dispersive X-ray spectroscopy (EDS, Oxford X-Max 50). The single crystal X-ray diffraction (XRD) patterns and room-temperature powder XRD pattern were carried out by a Brucker D8 Advance X-ray diffractometer using Cu K$_{\alpha}$ radiation. The magnetization measurements were performed on a Quantum Design magnetic properties measurement system (QD MPMS-3). The electrical transport measurements were measured using a Quantum Design physical property measurement system (QD PPMS-14 T). For angular dependent MR measurements, the samples were mounted on a rotation holder.

\section{\uppercase\expandafter{\romannumeral3}. Crystal structure and resistivity}

EuAuAs crystallizes in a hexagonal symmetry with the space group $P$6$_{3}$/mmc (No.194) [Fig. 1(a)]. The magnetic Eu layers alternate with the nonmagnetic Au–As layers along the $c$-axis, which constitute two-dimensional triangular and honeycomb lattices, respectively. The room-temperature powder XRD pattern and the corresponding Rietveld refinement are shown in Fig. 1(b). The obtained lattice parameters are $a$ = $b$ = 4.4381(7) \AA{} and $c$ = 8.2798(3) \AA{}, which are in good agreement
with previous report \citep{PhysRevB.105.045103}. The small reliability parameters R$wp$ = 4.74\% and R$p$ = 3.3\% suggest that the structural model is consistent with the diffraction data.
As shown in Figs. 1(c) and 1(d), the room-temperature single crystal XRD patterns exhibit only the (0l0) and (00l) Bragg peaks, respectively. The photographs of the corresponding single crystals are displayed in the insets. Figure 1(e) shows the temperature-dependent longitudinal resistivity $\rho_{xx}$ and $\rho_{zz}$ at zero field. As the temeperature decreases, the $\rho_{xx}$ and $\rho_{zz}$ gradually decrease, exhibiting the metallic behaviors. When temeprature approaches to 38 K, both $\rho_{xx}$ and $\rho_{zz}$ show an increase until near $T_{N}$, and then rapidly decrease [the inset of Fig. 1(e)]. The markedly larger $\rho_{zz}$ compared with $\rho_{xx}$ reveals a strong electronic anisotropy in the longitudinal resistivity.

\section{\uppercase\expandafter{\romannumeral4}. Magnetic properties } 

Field-cooled (FC) and zero-field-cooled (ZFC) magnetic susceptibility ($\chi$) versus temperature (T) were measured with $H$ = 0.1 T along the $c$-axis ($\chi_{c}$, blue line) and within the $ab$-plane ($\chi_{ab}$, red line) [Fig. 2(a)]. As shown the inset of Fig. 2(a), both $\chi_{ab}$ and $\chi_c$ exhibit a pronounced peak at approximately 5.7 K and 6.3 K, indicating a magnetic transition from the paramagnetic (PM) state to the AFM state. This transition temperature is consistent with the anomaly observed in $\rho(T)$ [Fig. 1(e)]. As shown in Fig. 2(b) and Fig. 2(c), the inverse ZFC magnetic susceptibility $1/\chi$($T$) curves can be fitted using modified Curie-Weiss (CW) law \cite{10.1093/oso/9780198505921.001.0001} $\chi$ = $\chi_{0}$ + $C/(T-\theta_{\text{CW}})$ in the temperature range of 50–300 K, where $\chi_{0}$ is the temperature-independent term, $C$ is the Curie constant, and $\theta_{\text{CW}}$ represents the Weiss temperature. 
The red lines correspond to the fitting curves of 1/$\chi_{ab}$ and 1/$\chi_{c}$. The Curie–Weiss temperatures are determined to be $\theta_{\text{CW}}^{ab}$ = 5.3 K and $\theta_{\text{CW}}^{c}$ = 3.3 K for $H \parallel ab$ and $H \parallel c$, respectively. The positive Weiss temperatures suggest ferromagnetic interaction in the paramagnetic regime. The effective magnetic moments are calculated by the relation $\mu_{\text{eff}}$ = $\sqrt{8C}$ $\mu_{\text{B}}$ \citep{10.1038/s42005-022-00853-y}. The $\mu^{ab}_{\text{eff}}$ = 7.74 $\mu_{\text{B}}$/Eu for $H \parallel ab$ and $\mu^{c}_{\text{eff}}$ = 8.19 $\mu_{\text{B}}$/Eu for $H \parallel c$, both close to the theoretical value of $g\sqrt{S(S+1)}$$\mu_{\text{B}}$ = 7.94 $\mu_{\text{B}}$/Eu for $g$ = 2 and $S$ = 7/2. Table II shows a summary of the extracted Curie-Weiss fitting parameters. Specifically, the effective magnetic moment $\mu^{c}_{\text{eff}}$ = 8.19 $\mu_{\text{B}}$/Eu is larger than the Eu$^{\text{2+}}$ ion theoretical value, the similiar phenomenon has also been observed in EuZn$_{2}$As$_{2}$ and EuCd$_{2}$As$_{2}$ \citep{PhysRevB.105.165122}.

The field-dependent magnetization curves $M(H)$ from 2 K to 50 K for $H \parallel ab$ and $H \parallel c$ are presented in Fig. 2(d) and Fig. 2(e). Below $T_{\text{N}}$, the $M(H)$ curves initially increase nonlinearly with $\mu_{\text{0}}H$ before saturating at high fields for $H \parallel ab$ and $H$ // $c$. Even above $T_{\text{N}}$ (e.g., at 30 K), the magnetization also exhibits a nonlinear field dependence, indicating the presence of short-range magnetic interaction. At 50 K (well above $T_{\text{N}}$), the nearly linear field-dependent magnetization is consistent with a paramagnetic state. The difference is that the $M(H)$ curves exhibit a small hysteresis in the field range of 0.1 T $<$ $ \mu_{0}H$ $<$ 0.6 T for $H \parallel ab$ [the inset of Fig. 2(d)]. In contrast, the magnetic hysteresis is not observed for $H \parallel c$ [the inset of Fig. 2(e)]. Moreover, Fig. 2(f) presents the $M(H)$ curves for $H \parallel ab$ and $H \parallel c$ measured from 0 to 7 T at 2 K, in which the saturation magnetization reaches 6.5 $\mu_{\text{B}}$/Eu at about 2.5 T for $H \parallel ab$ and 6.8 $\mu_{\text{B}}$/Eu at about 3.5 T for $H \parallel c$, both of which are close to the theoretical value $gS\mu_{\text{B}}$ = 7.0 $\mu_{\text{B}}$/Eu. This suggests that the spin moments are almost fully aligned with the external field at 7 T and that the easy magnetization direction lies in $ab$ plane. It is worth noting that the $M(H)$ curve shows a minor change in slope below approximately 1 T at 2 K for $H \parallel ab$. To visualize this feature more clearly, we plot the derivative $dM$/$dH$ versus $H$ shown in the inset of Fig. 2(f). For $H \parallel ab$, the $dM$/$dH$ curve exhibits a pronounced peak near $H_{ab}^{\text{t}}$ = 0.5 T, which corresponds to a metamagnetic transition. In contrast, the response for $H \parallel c$ remains smooth over the field range. These results are consistent with an easy-plane antiferromagnetic state, possibly with Eu$^{\text{2+}}$ magnetic moments ferromagnetically aligned within the $ab$ plane and antiferromagnetically coupled along the $c$ axis. The overall magnetic features are consistent with those reported isomorphic compounds such as EuAgAs \citep{PhysRevB.103.L241112}, EuCuAs \citep{j.jallcom.2014.02.157,10.1021/jacs.3c04249},  EuCuSb \citep{PhysRevB.102.174425,PhysRevB.111.054428}, $etc$.

\begin{figure*}[htpb]
	\centering
	\includegraphics[width=1\textwidth]{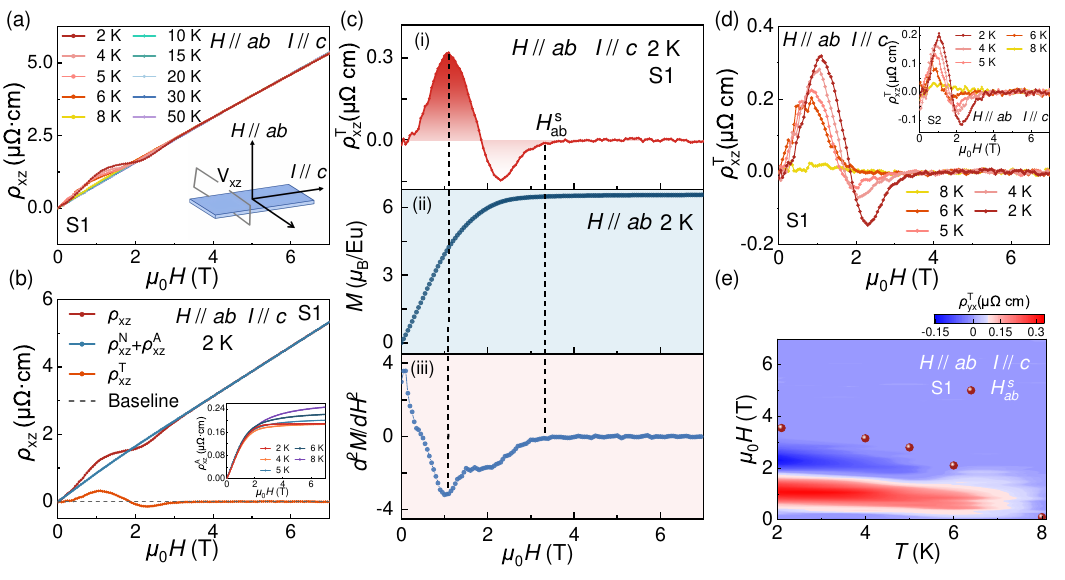}
	\caption{THE in EuAuAs. (a) Magnetic-field dependent Hall resistivity $\rho_{xz}$ measured at various temperatures with $H \parallel ab$ and $I \parallel c$. Inset is the schematic illustration of the measurement geometry. (b) 
	Hall resistivity analysis at 2 K. The red line shows the total Hall resistivity $\rho_{xz}$, the blue solid line represents the sum of the normal and anomalous Hall contributions, $\rho^{\text{N}}_{xz} + \rho^{\text{A}}_{xz}$ ($\rho^{\text{N}}_{xz}$ = $\mu_0 R_0 H$ and $\rho^{\text{A}}_{xz}$ = $\mu_0R_{S}M$). The extracted topological Hall resistivity, $\rho^{\text{T}}_{xz}$ = $\rho_{xz}$ - $\rho^{\text{N}}_{xz}$ - $\rho^{\text{A}}_{xz}$, is shown by orange solid line. The gray dashed line denotes $\rho^{\text{T}}_{xz}$ = 0. The inset shows the field-dependent $\rho^{\text{A}}_{xz}$ from 2 K-8 K. (c) The field dependence of $\rho^{\text{T}}_{xz}$, $M$ and its corresponding second derivates $d^{2}M$/$dH^{2}$ with $H \parallel ab$ at 2 K. The $H^{\text{s}}_{ab}$ is saturation field for $H \parallel ab$. (d) Field-dependent $\rho^{\text{T}}_{{xz}}$ curves of sample S1 from 2 to 8 K. Inset: the field-dependent $\rho^{\text{T}}_{{xz}}$ of sample S2. (e) The color rendering of THE as function of temperature and magnetic field with $H \parallel ab$ and $I \parallel c$.}
	\label{Fig3}
\end{figure*}

\begin{table}[htbp]
	\centering
	\caption{The Curie-Weiss fitting parameters of EuAuAs for $H \parallel ab$ and $H \parallel c$. $C$, $\mu_{\text{eff}}$, $\theta_{\text{CW}}$ are the Curie constant, the effective magnetic moment, and the Weiss temperature, respectively.}
	\label{1}
	\renewcommand\arraystretch{1.4}
	\setlength{\tabcolsep}{2.5mm}{
		\begin{tabular}{ccccc}
			\hline \hline 
			& Direction & $C$ [emu K/(mol$\cdot$Oe)] & $\mu_{\text{eff}}$ ($\mu_{\text{B}}$/Eu) & $\theta_{\text{CW}}$ (K)\\
			\hline	
			&$H$ // $ab$ & 7.48 & 7.74 & 5.3 \\
			&$H$ // $c$ & 8.38 & 8.19 & 3.3 \\
			\hline 
	\end{tabular}}
\end{table}

\section{\uppercase\expandafter{\romannumeral5}. Hall effect }

To further investigate the transport properties of EuAuAs under magnetic field, the field-dependent Hall resistivity measurements are performed from 2 K to 50 K with $I \parallel c$ and $H \parallel ab$. Figure 3(a) presents the obtained Hall resistivity ($\rho_{xz}$) and the inset illustrates the measurement configuration. Remarkably, $\rho_{xz}$ exhibits a clear hump-like anomaly below $\sim$3 T during the magnetization process for $T < T_{\text{N}}$. The anomaly shifts slightly to lower fields with increasing temperature and gradually disappears above 6 K. However, the hump-like anomaly cannot be fully accounted for by the ordinary Hall contribution and a magnetization-scaled anomalous Hall contribution, suggesting the presence of an additional Hall component. Similar anomalies in $\rho_{xy}(H)$ have been observed in other magnetic topological materials, such as GdPtBi \citep{10.1038/nphys3831}, Gd$_{{2}}$PdSi$_{{3}}$ \citep{10.1126/science.aau0968}, and EuCd$_{{2}}$As$_{{2}}$ \citep{https://doi.org/10.1002/advs.202207121,PhysRevLett.126.076602}, where they were attributed to the THE arising from noncoplanar spin textures. To clarify the origin of the observed hump-like anomaly, we analyze the Hall resistivity by separating its different contributions. The total Hall resistivity in magnetic systems contains three independent contributions:

\begin{equation}\label{equ1}
	\centering
	\rho_{xz} =\rho^{\text{N}}_{xz} + \rho^{\text{A}}_{xz} + \rho^{\text{T}}_{xz} = \mu_{0} R_{\text{0}}H + \mu_{0} R_{S}M + \rho^{\text{T}}_{xz}
\end{equation}
where $\rho^{\text{N}}_{xz}$, $\rho^{\text{A}}_{xz}$ and $\rho^{\text{T}}_{xz}$ represent the normal, anomalous, and topological Hall resistivity, respectively. The normal resistivity $\rho^{\text{N}}_{xz}$ is defined as $\rho^{\text{N}}_{xz}$ = $\mu_{0}R_{\text{0}}H$, where $R_{\text{0}}$ is the normal Hall coefficient. The $\rho_{xz}$ exhibits a positive linear field dependence at high fields, indicating that charge transport in EuAuAs are dominated by hole-type carriers. The anomalous Hall resistivity $\rho^{\text{A}}_{xz}$ = $\mu_{\text{0}} R_{\text{S}}M$, where $R_{\text{S}}$ is the anomalous Hall coefficient. 
The $\rho^{\text{T}}_{xz}$ denotes topological Hall resistivity, which derives from a real-space Berry phase acquired by electrons upon moving through spatially varying spin textures that possess finite SSC \citep{PhysRevLett.93.096806,10.1038/s43246-022-00238-2,PhysRevB.39.11413}. We use the Hall data measured at 2 K to illustrate the extraction of the normal, anomalous, and topological contributions from the total Hall resistivity [Fig. 3(b)]. A linear fitting of $\rho_{xz}/\mu_0H$ versus $M/\mu_0H$ is performed in the field range 4 T $\leqslant \mu_0H \leqslant 7$ T, where the magnetization is nearly saturated and the topological Hall contribution is assumed to be negligible. The intercept and slope are determined to be $R_0 = 0.735~\mu\Omega\cdot\text{cm T}^{\text{-1}}$ and $R_S = 0.208~\mu\Omega\cdot\text{cm T}^{\text{-1}}$, respectively. Using these parameters, we calculate the the normal Hall resistivity $\rho_{xz}^{\text{N}} = \mu_0 R_0 H$ and the anomalous Hall resistivity $\rho_{xz}^{\text{A}} = \mu_0 R_S M$. The topological Hall resistivity is extracted as $\rho_{xz}^{\text{T}} = \rho_{xz} - \rho_{xz}^{\text{N}} - \rho_{xz}^{\text{A}}$. As shown in Fig. 3(b), the red line denotes the total measured Hall resistivity $\rho_{xz}$, the blue line represents the sum of the normal and anomalous Hall resistivity $\rho^{\text{N}}_{xz}$ + $\rho^{\text{A}}_{xz}$, the orange line corresponds to the topological Hall resistivity $\rho^{\text{T}}_{xz}$, and the dashed line is the baseline $\rho_{xz}$ = 0. Inset of Fig. 3(b) shows the field dependence of the extracted $\rho_{xz}^{\text{A}}$ over the temperature range from 2 K to 8 K. The $\rho_{xz}^{\text{A}}$ is $\sim$ 0.19 $\mu$$\Omega$ cm at 2 K, and gradually increases with increasing temperature.

\begin{figure*}[htbp]
	\centering
	\includegraphics[width=1\textwidth]{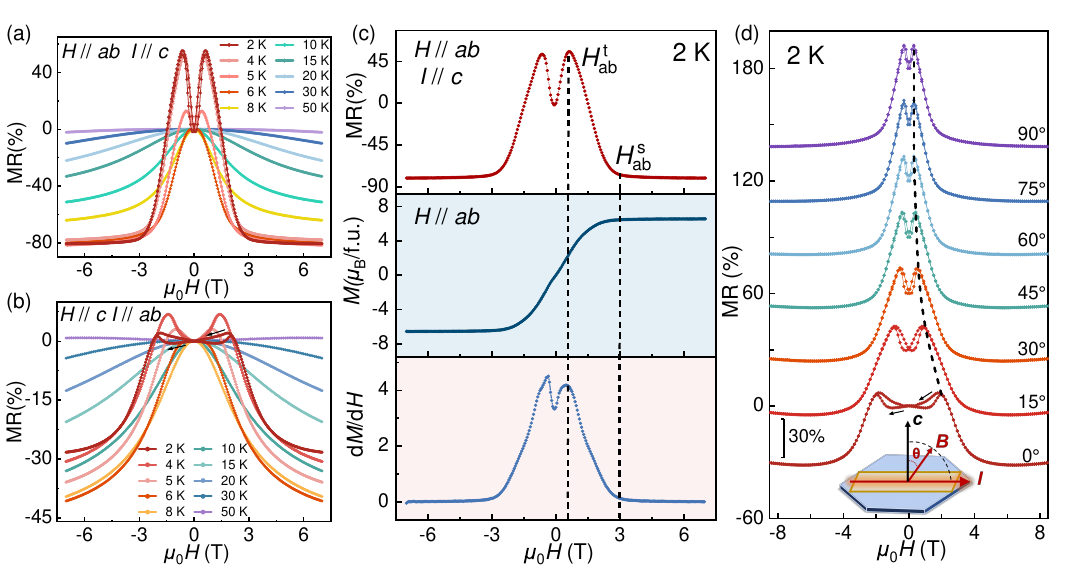}
	\caption{The magnetoresistance measurement of EuAuAs. (a) Magnetoresistance [MR = [$\rho_{xx}(H)-\rho_{xx}(0)$] / $\rho_{xx}(0)$$\times100\%$] as a function of $\mu_{0}H$ under $H \parallel ab$ and $I \parallel c$ at different temperatures. (b) MR as a function of $\mu_{0}H$ under $H \parallel c$ and $I \parallel ab$ at different temperatures. The black arrows represent the field-sweep direction from $+7$ T to $-7$ T. (c) The field- dependent MR, $M$ and the first derivates $dM$/$dT$ at $T$ = 2 K with $H \parallel ab$ and $I \parallel c$. The $H^{\text{t}}_{ab}$ and $H^{\text{s}}_{ab}$ are metamagnetic transition field and saturation field for $H \parallel ab$, respectively.
	(d) Field-dependent MR curves at different angles ($\theta$, defined in the inset) at 2 K. The butterfly-shaped MR is pronounced at $\theta$ = 0 and gradually weakens with increasing $\theta$.  Bottom inset shows the measurement geometry: the magnetic field $B$ is rotated from the $c$ axis toward the $ab$ plane with $I \parallel ab$.  The black arrows represent the field-sweep direction from $+7$ T to $-7$ T. The curves are vertically shifted upward by 30\% for clarity.}  
	\label{Figure} 	
\end{figure*}

To better understand the topological Hall response in EuAuAs, the field evolution of $\rho^{\text{T}}_{xz}$ and $M$, together with the second derivative $d^{2}M$/$dH^{2}$ at 2 K are summarized in Fig. 3(c). As shown in Fig. 3(c)(i), the magnitude of $\rho^{\text{T}}_{xz}$ reaches to 0.32 $\mu$$\Omega$ cm at 1.1 T, which is comparable to that of MnSi \citep{PhysRevLett.102.186602} and FeGe thin films \citep{PhysRevLett.118.027201}. Notably, $\rho^{\text{T}}_{xz}$ changes sign from 0.32 $\mu$$\Omega$ cm at 1.1 T to -0.15 $\mu$$\Omega$ cm at 2.2 T, and vanishes at the saturation field ($H^{s}_{ab}$) 3.35 T. Comparison among Fig. 3(c)(i)–(iii) reveal that the field-dependent evolution of $\rho^{\text{T}}_{xz}$, $M(H)$, and its second derivative are all consistent. 
The peak appears exclusively in the range 0 $<\mu_{0} H<H^{s}_{ab}$, suggesting the topological Hall response is associated with the magnetic-field-induced spin textures. The sign reversal of $\rho_{xz}^{\mathrm{T}}$ may reflect a field-induced reconstruction of the noncoplanar antiferromagnetic state. Similar relation between the topological Hall response and field-induced magnetic transformations have been reported in Fe$_5$Sn$_3$ \citep{10.1063/5.0005493} and ErMn$_6$Sn$_6$ \citep{PhysRevB.104.L161115}, with the latter also exhibiting a sign-changing THE near a metamagnetic transition. The sign reversal in EuAuAs may arise from the competition between anisotropic magnetic configurations that produce oppositely signed net SSC.

Figure 3(d) shows the field-dependent $\rho^{\text{T}}_{xz}$ curves of sample S1 at different temperatures. A pronounced peak-like anomaly appears in the low-field region and gradually weakens with increasing temperature, almost disappearing above 8 K (close to the Néel temperature $T_{\text{N}}$). Notably, a negative dip is observerd at intermediate fields and also exhibits a clear temperature dependence. With increasing temperature, this negative peak is progressively suppressed and vanishes at 6 K. To further clarify the field–temperature evolution of the topological Hall response, a ($T$,$B$) contour map for EuAuAs under $H \parallel ab$ and $I \parallel c$ was constructed from the temperature-dependent $M(H)$ curves and $\rho^{\text{T}}_{xz}$ data of S1 [Fig. 3(e)]. The topological Hall signal is pronounced only below $T_{\text{N}}$ and in the field region $H < H^{\text{s}}_{ab}$, indicating that it may be associated with the evolution of the magnetic structure. To verify whether this phenomenon is intrinsic to the materials, additional measurements are performed on sample S2. As shown in the inset of Fig. 3(d), sample S2 exhibits a similar hump-like anomaly and temperature dependence, which further support our experimental results. This topological Hall response is unlikely to originate from conventional skyrmions stabilized by a uniform bulk Dzyaloshinskii–Moriya interaction (DMI) \citep{PhysRevLett.102.186602,PhysRevLett.108.267201}, which is generally not expected in centrosymmetric systems. Another possible origin is thermal spin fluctuations above $T_{\mathrm{N}}$, which can generate spin clusters with finite SSC \citep{10.1002/adfm.202502016,10.1126/sciadv.aap9962}. Such fluctuation-induced chiral correlations will still persist above the magnetic transition temperature \citep{10.1002/adfm.202502016,10.1126/sciadv.aap9962}, whereas the THE signal disappears above AFM transition temperature in EuAuAs. The THE signal appears only in the antiferromagnetic state and vanishes above $H_{ab}^{\mathrm{s}}$. Combined with the metamagnetic transition for $H \parallel ab$, suggesting it may be associated with the field-induced evolution of the antiferromagnetic spin textures. One possible microscopic scenario is that this evolution stabilizes a noncoplanar spin configuration with a nonzero net scalar spin chirality ($\boldsymbol{\chi}_{ijk}\neq 0$). Conduction electrons moving through such spin textures can acquire a real-space nonzero Berry phase, thereby giving rise to a topological Hall response, as observed in Fe$_{1.3}$Sb \citep{PhysRevLett.108.056601}.

\section{\uppercase\expandafter{\romannumeral6}. MAGNETORESISTIVITY }

The effect of magnetic-field-induced spin textures on transport properties is also reflected on the MR, defined as MR = [$\rho_{xx}(H)-\rho_{xx}(0)$] / $\rho_{xx}(0)$$\times100\%$. Figure 4(a) presents the field-dependent MR curves with $H \parallel ab$ and $I \parallel c$ at different temperatures. At low temperatures and low magnetic fields, we can observe a weak butterfly-shaped negative MR, which is consistent with the small hysteresis in $M(H)$ curves [the inset of Fig. 2(d)]. Upon further increasing magnetic field, the MR monotonically decreases, which may be associated with the suppression of spin-disorder scattering as the system approaches the spin-polarized state. Similar MR behavior has also been reported in the antiferromagnet MnBi${_{2}}$Te${_{4}}$ \citep{PhysRevResearch.1.012011}. Meanwhile, as the temperature increases, the MR peak shifts to lower fields and gradually weakens, eventually disappears above $T_{\text{N}}$. At 6 K, the value of MR is approximately -81\%, which may arise from thermal spin fluctuations, together with the suppression of spin-dependent scattering under the applied field \citep{10.1016/j.jallcom.2022.165089}. The field evolution of the MR peak can be understood by comparing it with the magnetization behavior. As shown in Fig. 4(c), the MR peak is observed near the metamagnetic transition field $H_{ab}^{\text{t}}$ extracted from the $M(H)$ curves, indicating that the low-field MR anomaly may originate from the field-induced reorientation of the Eu$^{\text{2+}}$ moments. Above the saturation field $H_{ab}^{\text{s}}$, the MR exhibits a weak field dependence.

 For $H \parallel c$, the $M(H)$ curves evolve smoothly without an obvious metamagnetic transition, while the corresponding MR curves show a pronounced butterfly-shaped hysteresis and non-monotonic field dependence at low temperature [Fig. 4(b)]. The black arrow represents the field-sweep direction from $+7$ T to $-7$ T. The butterfly-shaped MR may be associated with the evolution of antiferromagnetic domains and domain-wall pinning during the field-induced spin reorientation. Similar butterfly-shaped MR hysteresis has also been reported in antiferromagnetic topological insulator MnBi$_{2}$Te$_{4}$ \citep{10.1038/s41563-019-0573-3}, CrI$_{3}$ \citep{10.1126/science.aar3617}, Cr$_{2}$Ge$_{2}$Te$_{6}$ \citep{10.1021/jacs.9b06929} and exfoliated flake of Cr$_{1/3}$TaS$_{2}$ \citep{10.1088/2053-1583/aa8a2b}.

To further clarify the relationship between MR and field orientation, angle-dependent MR measurements were performed by rotating the magnetic field from $H \parallel c$ to $H \parallel ab$. Figure 4(d) shows the magnetic-field-dependent MR for different angles $\theta$ at 2 K. The measurement configuration is shown in the inset, and the field sweep direction is plotted by black arrows. At $\theta$ = 0\degree, the MR exhibits a pronounced butterfly-shaped hysteresis. As the magnetic field rotates from the $c$ axis ($\theta$ = 0\degree) toward the $ab$ plane ($\theta$ = 90\degree), the hysteresis is gradually suppressed and almost disappears at $\theta$ = 90\degree. Meanwhile, the peak of the MR gradually shifts to a lower magnetic field, as marked by the black dashed line. This pronounced angular dependence suggests that the butterfly-shaped MR may be associated with the magnetic anisotropy, possibly involving the field-driven evolution of antiferromagnetic domains and domain-wall pinning \citep{10.1088/2053-1583/aa8a2b}.

\section{\uppercase\expandafter{\romannumeral7}. Summary  }

In this work, we synthesized single crystals of the centrosymmetric antiferromagnet EuAuAs and investigated the magnetic and electrical transport properties. Magnetic characterization reveals antiferromagnetic transitions at $T_{\text{N}}\approx5.7$ K and 6.3 K for $H \parallel ab$ and $H \parallel c$, respectively. The magnetization exhibits a weak metamagnetic transition accompanied by hysteretic behavior with $H \parallel ab$ below $T_{\text{N}}$. In the AFM state, a THE is observed with $H \parallel ab$ and $I \parallel c$, which is likely associated with nonzero scalar spin chirality arising from a field-induced noncoplanar spin structure. Furthermore, a distinct butterfly-shaped magnetoresistance is observed for $H \parallel c$ and $I \parallel ab$. The strong angular dependence of MR reveals a correlation between the butterfly-shaped MR hysteresis and the magnetic anisotropy of the system, possibly involving the field-driven evolution of antiferromagnetic domains and domain-wall pinning. These observations suggest the field-induced spin textures and magnetic -domain dynamics play an important role in the magnetotransport properties in EuAuAs.

\section{Acknowledgments}

This work is supported by the National Key R\&D Program of China (Grant No. 2024YFA1409002), the National
Natural Science Foundation of China (Grant No. 12404191), the Fundamental Research Funds for
the Central Universities, and the Research Funds of Renmin
University of China (Grant No. 23XNKJ22), and the Beijing
National Laboratory for Condensed Matter Physics (Grant
No. 2023BNLCMPKF008).

\bibliography{Bibtex}
\end{document}